\documentclass[aps,nofootinbib,reprint,nobibnotes,showpacs,superscriptaddress,floatfix]{revtex4-1}
\usepackage[english]{babel}
	\addto\captionsenglish{%
	}

\usepackage{graphicx}
\usepackage{subfigure}
\usepackage{multirow}
\usepackage{amsmath}
\usepackage{amssymb}
\usepackage[version=3]{mhchem}	% chemical formulas -> e.g. \ce{^{130}Te} instead of $^{130}$Te
\usepackage{bm}	% bold
\usepackage{nicefrac}	% fractions: a/b
\usepackage{mathrsfs}
\usepackage[normalem]{ulem}    % to striketrhourhg text

\usepackage[usenames,dvipsnames,svgnames,table]{xcolor}	%colored text
\usepackage{comment}	% commented (non considered) enviroment \begin{comment} ... \end{comment}
\usepackage{upgreek}	% non italic ÎŒ symbol

\usepackage{textcomp}	% copyright logo

\newcommand{\bb}{$0\nu \beta \beta$} % neutrinoless double beta decay
\newcommand{\kg}{\text{kg}}		

		\newcommand{\K}{\text{K}}

		\newcommand{\W}{\text{W}}

\newcommand{\hyt}[1]{\hypertarget{#1}{#1}}					% intra-test hypertarget
\begin{document}

\title{An active noise cancellation technique for the CUORE Pulse Tube Cryocoolers}

	\author{A.\ D'Addabbo}
		\email[Corresponding author - ]{antonio.daddabbo@lngs.infn.it}
	\author{C.\ Bucci}
		\affiliation{INFN -- Laboratori Nazionali del Gran Sasso, Assergi (L'Aquila) I-67100, Italy}
	\author{L.\ Canonica}
		\thanks{Currently at Max-Planck-Institut f\"ur Physik, D-80805, M\"unchen, Germany}
		\affiliation{INFN -- Laboratori Nazionali del Gran Sasso, Assergi (L'Aquila) I-67100, Italy}
		\affiliation{Massachusetts Institute of Technology, Cambridge, MA 02139, USA}
	\author{S.\ Di Domizio}
		\affiliation{INFN -- Sezione di Genova, Genova I-16146, Italy}
		\affiliation{Dipartimento di Fisica, Universit\`a di Genova, Genova I-16146, Italy}
	\author{P.\ Gorla}
		\affiliation{INFN -- Laboratori Nazionali del Gran Sasso, Assergi (L'Aquila) I-67100, Italy}
	\author{L.\ Marini}
		\thanks{Currently at Department of Physics, University of California, Berkeley, California 94720, USA}
		\affiliation{INFN -- Sezione di Genova, Genova I-16146, Italy}
        \affiliation{Dipartimento di Fisica, Universit\`a di Genova, Genova I-16146, Italy}
    \author{A.\ Nucciotti}
		\affiliation{INFN -- Sezione  di  Milano  Bicocca, Milano  I-20126, Italy}
        \affiliation{Dipartimento  di  Fisica,  Universit\`a  di  Milano-Bicocca,  Milano  I-20126,  Italy}
    \author{I.\ Nutini}
		\affiliation{INFN -- Laboratori Nazionali del Gran Sasso, Assergi (L'Aquila) I-67100, Italy}
        \affiliation{Gran Sasso Science Institute -- L'Aquila, I-67100, Italy}
    \author{C.\ Rusconi}
    	\affiliation{INFN -- Laboratori Nazionali del Gran Sasso, Assergi (L'Aquila) I-67100, Italy}
		\affiliation{Department of Physics and Astronomy, University of South Carolina, Columbia, SC 29208, USA}	
	\author{B.\ Welliver}
		\affiliation{Nuclear Science Division, Lawrence Berkeley National Laboratory, Berkeley, CA 94720, USA}	

\quad

\date{\today}

\begin{abstract}
		
The Cryogenic Underground Observatory for Rare Events (CUORE) experiment at Gran Sasso National Laboratory of INFN searches for neutrinoless double beta decay using TeO$_2$ crystals as cryogenic bolometers. The sensitivity of the measurement heavily depends on the energy resolution of the detector, therefore the success of the experiment stands on the capability to provide an extremely low noise environment. One of the most relevant sources of noise are the mechanical vibrations induced by the five Pulse Tube cryocoolers used on the cryogenic system which houses the detectors.
To address this problem, we developed a system to control the relative phases of the pulse tube pressure oscillations, in order to achieve coherent superposition of the mechanical vibrations transmitted to the detectors. In the following, we describe this method and report on the results in applying it to the CUORE system.
\end{abstract}	

\maketitle

\section{Introduction} \label{sec:intro}
In the last two decades rare event searches have sharply accelerated. The determination of the nature of the neutrino and the search for dark matter are crucial for our understanding of new physics beyond both the well established Standard Model of particle physics \cite{Herrero:1998eq} and the Lambda Cold Dark Matter cosmological model \cite{Ade:2015xua}. Although the experimental techniques substantially differ, these searches share the common need for large mass detectors. Indeed the only way to maximize the probability of detection is to increase the number of source/target nuclei that are under investigation.
	
Among the available detection techniques, the bolometric approach offers an intrinsically high energy resolution, but suffers from a limited scalability to high masses due to cryogenic constraints. In order to provide the necessary cooling power at millikelvin temperatures for a ton-scale detector, a custom cryogenic apparatus exploiting dilution refrigeration is needed. The dilution process requires a 4\,K environment which can be maintained  by using either a liquid He (LHe) bath or mechanical cryocoolers, such as Pulse Tubes (PTs) \cite{EnssSiegfried:2005}. Although the use LHe has proved to be a limiting factor in the scalability of cryogenic experiments -- both in terms of cost and duty cycle -- in past years it has been preferred over the cryogen-free approach due to burden of mechanical vibrations caused by cryocoolers.
	
In the following we describe an innovative technique which has been developed for the CUORE experiment to  drastically suppress the noise induced by the PTs. The implementation of this technique will ease the development of large cryogen-free systems for future rare events searches. First, we introduce the CUORE experiment (\ref{sec:CUORE}) and its PT cryocoolers (\ref{sec:CUORE_PTs}). Next, we illustrate the technique developed (\ref{sec:drivePT}) and describe the analysis method (\ref{sec:anc}). Finally, we report the results obtained for the CUORE system (\ref{sec:results}). 

\section{The CUORE challenge} \label{sec:CUORE}
	CUORE is a ton-scale experiment that searches for neutrinoless double beta decay of $^{130}$Te (\bb, ~\cite{Artusa:2014lgv}). This is accomplished using TeO$_2$ crystals operating as cryogenic bolometers with a total mass of 742 $\kg$ ($\sim 206\,\kg$ of \ce{^{130}Te}). The detector consists of 988 crystals arranged into 19 identical structures called ``towers''. Each tower hosts 52 bolometers arranged in 13 floors.
	
	The CUORE cryostat has been designed to house this ton-scale detector and to maintain it at about 10 mK continuously for 5 to 10 years. To face this unprecedented challenge, a large custom cryogen-free cryostat has been designed \cite{CUORECryostat}. Five PT cryocoolers cool down to 35\,K and 3.5\,K respectively 980\,kg and 7400\,kg of material, mainly copper plates and vessels and $^{210}$Pb-depleted archaeological lead shielding \cite{Alessandrello:1998RL}. A high-power (3 $\mu$W at 12 mK) \ce{^3He}/\ce{^4He} Dilution Unit (\hyt{DU}) then provides cooling power to cool to around 10 mK the 1519 kg of the detector structure with its shielding vessel  - mostly copper and TeO$_2$ crystals.

\begin{comment}		
	\begin{figure*}[t]
		\centering
		\includegraphics[width=.8\textwidth]{cryo_nucc_1_02_ss}
		\caption{\FIG Schematic of the CUORE cryostat.
			\lam{Immagine tipo questa, con tutte le strutture be visibili (con frecce e nomi) e in doppia colonna}}
		\label{fig:cryostat_schematic}
	\end{figure*}
\end{comment}

	In order to minimize the detector noise, the entire experimental volume -- about one cubic meter -- must be kept in optimal and stable experimental conditions, which means in an exceptionally low radioactive background and a low mechanical vibration environment. To address the former requirement, all the materials used in CUORE meet strict radio-purity criteria and underwent different types of cleaning procedures (\cite{Alduino:2016vjd} and references therein). To cope with the latter requirement, a special suspension has been designed to mechanically decouple the detector structure from the rest of the cryostat. The suspension insulator has been custom designed from Minus-K technology and acts as a mechanical low pass filter for vibrations, with a cut-off frequency of about 0.5\,Hz in both the vertical and horizontal directions \cite{CUORESuspension}. Moreover, the whole CUORE infrastructure itself has been designed to be isolated from external sources of vibrations by means of four elastomeric dampers.	

\section{The CUORE PTs} \label{sec:CUORE_PTs}
	\begin{figure}[t!]
		\centering
		\includegraphics[width=0.9\columnwidth]{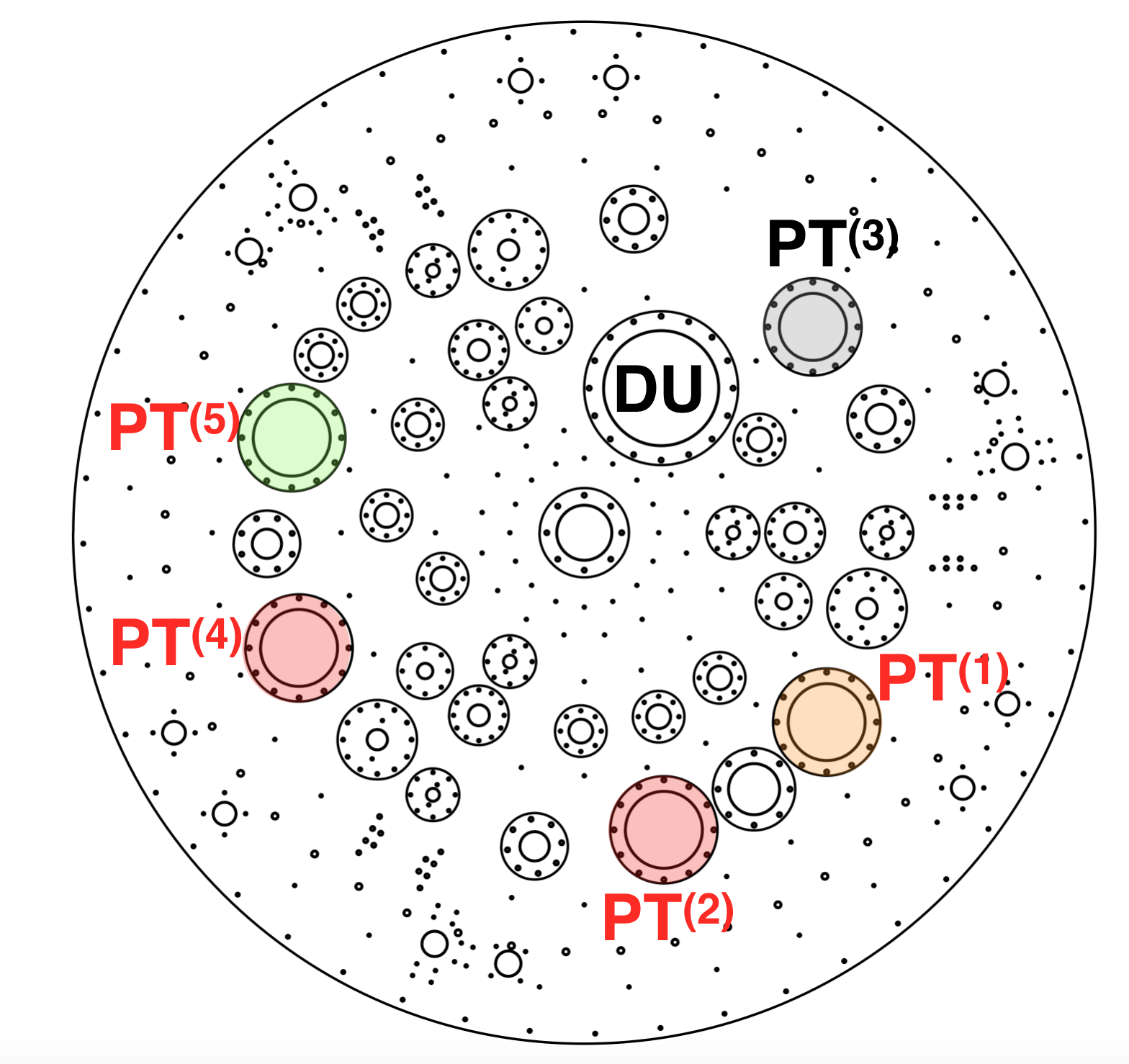} \\[+10pt]
		\caption{Top view of the 300 K plate of the CUORE cryostat. The positions of the five PTs and of the Dilution Unit (DU) are highlighted. PT$^{(3)}$ is not powered during data taking. Colors refer to optimal phase configuration commented in sec. \ref{sec:results}: PT$^{(4)}$ and PT$^{(2)}$ are in phase opposition (red) with respect to the reference PT$^{(5)}$ (green), while PT$^{(1)}$ is in quadrature (orange) with respect to PT$^{(2)}$.}
	\label{fig:300K}
	\end{figure}
	
	A Pulse Tube is a cryocooler whose cooling power is provided by means of $^4$He gas isoenthalpic expansions. The cooling effect relies on a periodic variation of the pressure inside one or more thin-walled tubes -- the actual ``Pulse Tubes'' -- containing a large heat capacity regenerator and with heat exchangers at both ends~\cite{Pobell:2007}.

	The pressure cycles are provided by a rotary valve inside the PT motor-head at room temperature. This valve makes $0.7$ revolutions per second and alternatively connects the PT to the high-pressure and low-pressure side of a compressor. This results in pressure waves with a frequency that depends on the PT model and which is 1.4\,Hz for the CUORE PTs.
	
	In cryogenic systems such as the ones used for rare event searches, PT cooling is becoming more popular compared to LHe bath -- even with the support of systems to reduce boil-off or for in-situ reliquefaction -- for several reasons, which include cost, reliability, stability, duty cycle, and safety.
	
	The downside of mechanical cryocoolers is normally the production of mechanical vibration inside the experimental apparatus. With PTs the absence of moving parts at low temperature increases the reliability and strongly reduces the magnetic interference and the intensity of vibrations generated during operation with respect to other cryocoolers, such as piston-based Gifford-McMahon heads. Small residual mechanical vibrations still occur in a PT due to the elastic deformation caused by pressure oscillations in the He gas inside them.
	
	All these features make the PTs suitable for cooling sensitive devices like the low temperature detectors used for CUORE, but care must be taken to minimize the impact of their residual  mechanical vibrations.
	
	The large number of PTs installed  in CUORE cryogenic system is determined by the significant heat load in this large system which derives from thermal radiation, conduction along mechanical supports, nearly 3000 electrical wires, and $^3$He circulation for the dilution unit.
	
	To provide the needed cooling power in all foreseeable running conditions -- with a safe margin -- CUORE mounts five PT415-RM by Cryomech \cite{PT415-RM}. For CUORE, the Remote Motor option flexline is longer than the standard RM product offered by Cryomech (2\,ft long instead of 1\,ft). This solution offers a more efficient mechanical decoupling at the cost of a $\sim$ 10\% reduction in the cooling power with respect to the standard PT415-RM model. With this feature, the nominal cooling power of each PT is $1.2\,\W$ at $4.2\,\K$ and $32\,\W$ at $45\,\K$ each, while the lowest temperature achievable is close to $3\,\K$ with no thermal load \cite{Delloro_PhD-thesis:2017}. 
With these performances, the cooling power provided by no more than 4 PTs is sufficient to operate the CUORE dilution unit with the maximum cooling power. The fifth PT is therefore kept as a spare. The CUORE PTs are arranged (see Fig. \ref{fig:300K}.) on a circumference with a radius of $\sim$ 2/3 of the top cryostat plate (300 K plate) in a non symmetric way. PT$^{(1)}$ and PT$^{(5)}$ are the most critical as the DU condensing lines (which carries the $^3$He/$^4$He gas from room temperature to the DU) are thermalized on them. When running the cryostat with only four PTs, any other of the three PTs can be switched off. We keep PT$^{(4)}$ and PT$^{(2)}$ active for this analysis.

\subsection{Vibration noise}
\label{sec:vib}

	The effect of mechanical vibration on the CUORE detectors is two-fold: electrical and thermal.
The coupling of vibrations to the signal wires, which run from the detectors to the room temperature connectors, causes noise due to parasitic capacitance ($\sim$500\,pF/m) combined with the high impedance of the detector thermal sensors (order 200\,M$\Omega$). Despite a differential read-out scheme \cite{Arnaboldi:2017aek}, these cause an electrical microphonic noise pickup which can diminish the S/N of the detectors.

	Mechanical vibration transmitted by the PT head to the 300\,K flange and from the two cold stages of the PT to the 40\,K and 4\,K flanges, can propagate to the coldest parts of the system generating an excess heat load. The most affected parts are the DU stage with the lowest cooling power, i.e. the Mixing Chamber, and the weakly thermally coupled TeO$_2$ crystals.

	\begin{figure}[t]
		\centering
		\includegraphics[width=1.\columnwidth]{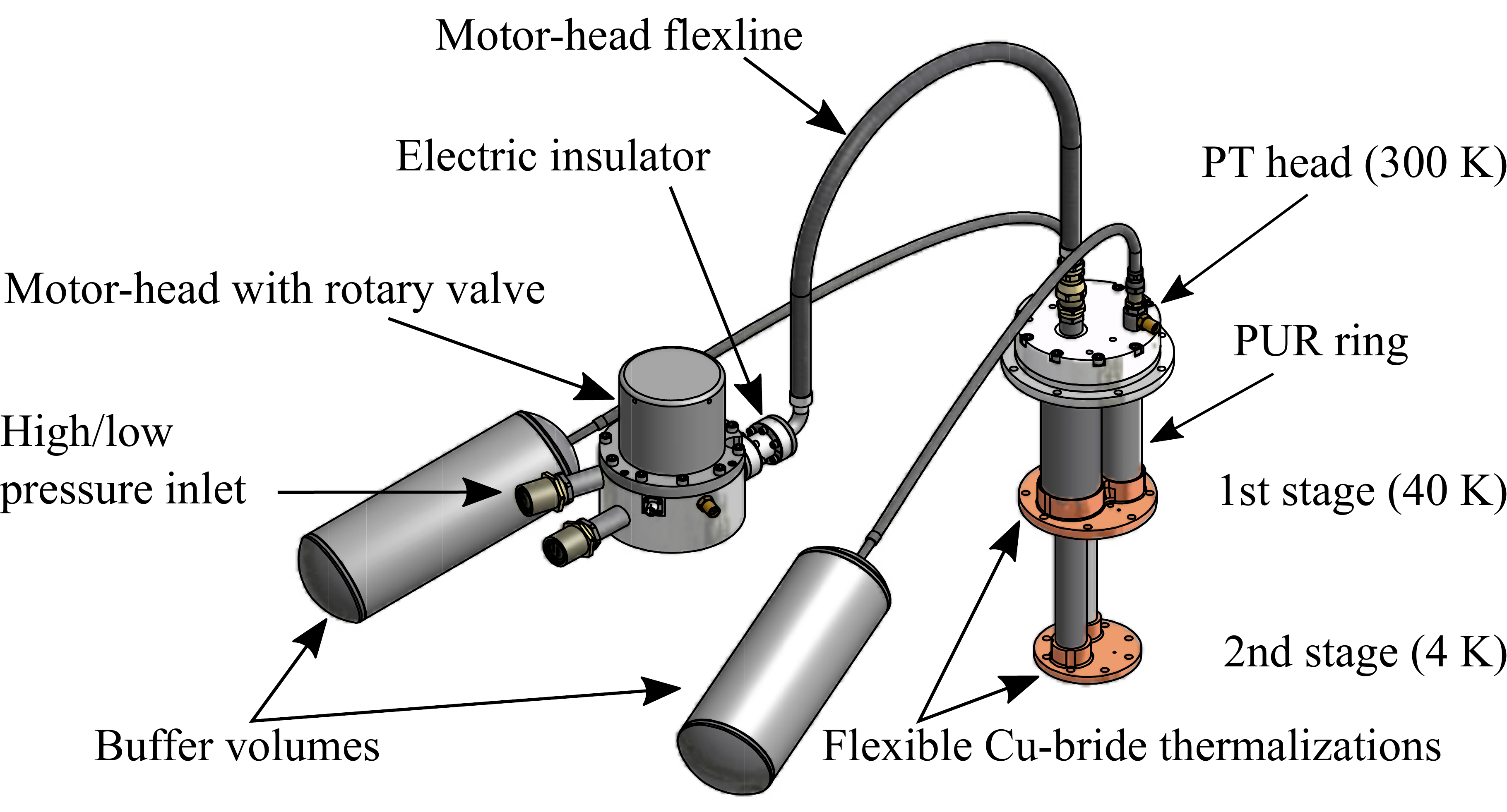}
		\caption{Rendering of a CUORE PT.}
		\label{fig:PT}
	\end{figure}

\subsection{Vibration control}
\label{sec:PT_dec}

	Vibrations induced on the coldest stages of cryogen-free cryostats are certainly a well known problem to those operating dry dilution refrigerators. In order to mitigate the vibrations induced by the PT cryocoolers, several passive solutions have been proposed to date \cite{Olivieri:2017lqz}. In CUORE, a series of passive ``standard" counter measures are enacted to minimize the vibrational noise due to coupling from the PTs into the detector system.
	
	PTs generate mechanical vibrations mainly in 3 ways. The first source of vibration is the only moving part of these devices: the rotary valve. The valve is moved by a DC stepper motor to which is mechanically coupled. The motor in turn is driven at the selected rotation speed by a microstepping driver embedded in the CP1010 Cryomech compressor. The gas pulsation generated by the motor and valve rotation is the source of vibration at 1.4 Hz and related harmonics. Several studies have shown how decoupling the PT heads and rotary valves from the dilution refrigerators are essential for reducing noise vibrations \cite{Ikushima:2008}. Such a solution has been adopted for each CUORE PT, whose rotary valve is separated from its head by means of a flexible Stainless steel line (1/2" ID, 2ft long) (see Fig. \ref{fig:PT}). In the following, they will be referred to as ``motor-head flexlines''. A combination of custom made rigid lines and commercial flexlines connects the compressor to its respective motor-head. In the following, they will be referred to as ``compressor flexlines''. They provide the standard high-pressure side and low-pressure side of the helium that reaches the motor-heads, and their path carefully avoids any contact with the cryostat. The gas high/low pressure compressor lines, the rotary valves, and the buffer volumes are suspended from the room ceiling by means of elastic bands to further absorb the vibrations. For the same reason, a swan-neck geometry is adopted for the motor-head flexlines. 

\begin{figure}[!t]
		\centering
		\includegraphics[width=0.6\columnwidth]{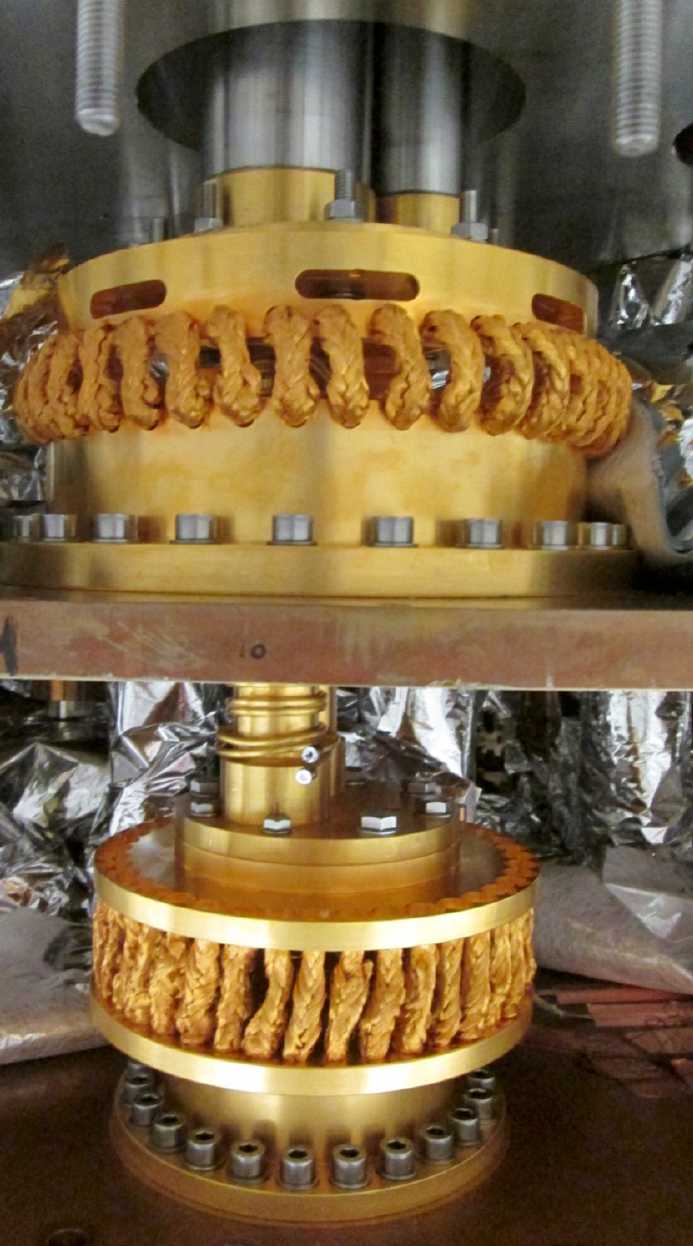}
		\caption{Picture of the thermal link between the cold PT head and the CUORE plates at the two temperature stages of 40\,K and 4\,K.}
		\label{fig:PTbraids}
	\end{figure}

The other vibration sources are related to the He pressure waves. These cause 1) periodic stress forces between the rotary valve and the PT head, and 2) periodic elastic deformations of the two coupled thin walled tubes of the two cooling stages.
The resulting effects of this pressure wave are a periodic force applied to the mounting point at room temperature and a dynamic displacement of the cold stages. 
For example Cryomech has measured the 2nd stage maximum displacement to be around 10 $\mu$m and 20 $\mu$m in the horizontal plane and in the vertical direction, respectively \footnote{From Cryomech private communications.}. The horizontal displacement is almost circular because of the asymmetric designs of the PT.

The PTs are mounted on the 300K flange of the cryostat by means of a sliding-seal arrangement. To reduce the vibration transmission to the cryostat flange through the mechanical coupling, leak tightness is provided by a specially selected soft o-ring. To further absorb and dissipate the mounting forces, a Polyurethane (PUR) disk is sandwiched between the PT head and the cryostat flange. In the innermost portion of the cryostat, the PT cold stages inject vibrations through their unavoidable thermal links to the 40\,K and 4\,K flanges. These thermal links have been designed to minimize the mechanical coupling while providing a strong enough thermal path. They are made by soft ETP 5649/71 copper braids with their ends TIG welded to copper blocks bolted on the PT stages and on the cryostat flanges. The copper braid is made up of several 0.07\,mm thin copper wires braided into tubes with a total effective cross section of 20 mm$^2$. The thermal links are designed with the aim to keep the thermal gradients between the PT stages and the cryostat flanges smaller than about 5\%. The thermal links for the two PT stages (Fig. \ref{fig:PTbraids}) use 40 braids 63\,mm long and 36 braids 60\,mm long to provide thermal conductions of about 20\,W/K at 40\,K and 7.5\,W/K at 4\,K, respectively.

\section{Active noise cancellation} \label{sec:drivePT}
	Despite of all the precautions taken to ensure a quiet environment for the detectors, the combined simultaneous operation of several PTs at the same 1.4\,Hz frequency has the potential to disrupt the detector performance. A non-negligible fraction of vibrational noise at the main frequency and its harmonics leaks down to the coldest part of the cryogenic system. The observed effects are mostly on the temperature of the Mixing Chamber (MC) and on the microphonic noise picked up by the read-out wires. Thanks to the effectiveness of the suspension system, the thermal impact on the detectors should be suppressed. Nonetheless, a non-negligible portion of the mechanical energy carried by the oscillations still reaches the detector. The work presented in this paper outlines a novel active vibration reduction designed for multiple PT cryocoolers in a dry dilution refrigerator setup.
	
	The technique that we propose has the purpose to control the relative phases of the CUORE PTs in order to find the configuration that maximizes the noise cancellation leveraging the interference between the noise sources. Given the complexity and the asymmetry of the system (see Fig. \ref{fig:300K}), the optimal configuration has been obtained scanning a large number of possible phase configurations to achieve the lowest detector noise. This aim has been accomplished in two steps. First, the Cryomech stepper motor drives for the rotary valve have been replaced by new low-noise, linear stepper motor drives in order to have more precise and stable motion control. Second, a complete scan of the relative phases among the four active PTs has been performed to identify the noise minima and tune the system to the most convenient one. The noise level on the coldest stage of the cryostat has been evaluated by acquiring the noise power spectra on the CUORE detector itself, with each of the 988 crystals operated as a cryogenic bolometer. On each bolometer, the noise is measured via a Ge-NTD directly glued on the surface of the crystal. This in turn provides us with 988 noise readings. For a detailed description of the CUORE bolometers refer to \cite{Alduino:2016vjd}.  

\subsection{Vibrational Beats}

	Due to the fact that the driving frequencies of the PTs are not exactly the same, their relative phases and the interference pattern from vibrations will vary with time.
	
	Beats will be generated between any pair of slightly different PT working frequencies, $f$ and $f + \Delta f$. The main frequency, $f_{mod} = f + \Delta f / 2$ shows up with the amplitude  $2 \cos{2\pi f_{beat}}$, which is modulated by the beat frequency
\begin{equation} \label{eq:0.0}
f_{beat} = \Delta f / 2
\end{equation}
Here $\Delta f$ can represent the intrinsic inaccuracy in writing the PT frequency.

\subsection{Linear Drive}

	The PT415-RM rotary valve is normally driven by a board embedded in the Cryomech compressor which is not optimized for low noise operation and which is not ready for user remote control. 
CUORE instead uses a LNX Linear Series low-noise drive from Precision Motion Control. It is a stepper motor device with a linear amplifier for the output stage, that we will refer to as Linear Drive (LD) in the following. It is intended for micro-stepping precision and customized linear and rotary motion solutions where motor vibrations must be minimized. This is achieved by smoothing the rotation of the PT valve, dividing the 360$^{\circ}$ angle into 25600\, possible positions (step numbers). Besides allowing precise control of the rotation frequency, the LD also reduces the vibrations produced by the rotary valve. The frequency control is an essential feature in order to be able to modify the relative phases of the pressure waves generated by the rotary valves. To measure and monitor these phases, all the low-pressure sides of the compressor flexlines have been equipped with pressure sensors, which are digitized and sampled at 2\,kHz.
   
   \begin{figure}[t]
		\centering
		\includegraphics[width=1.\columnwidth]{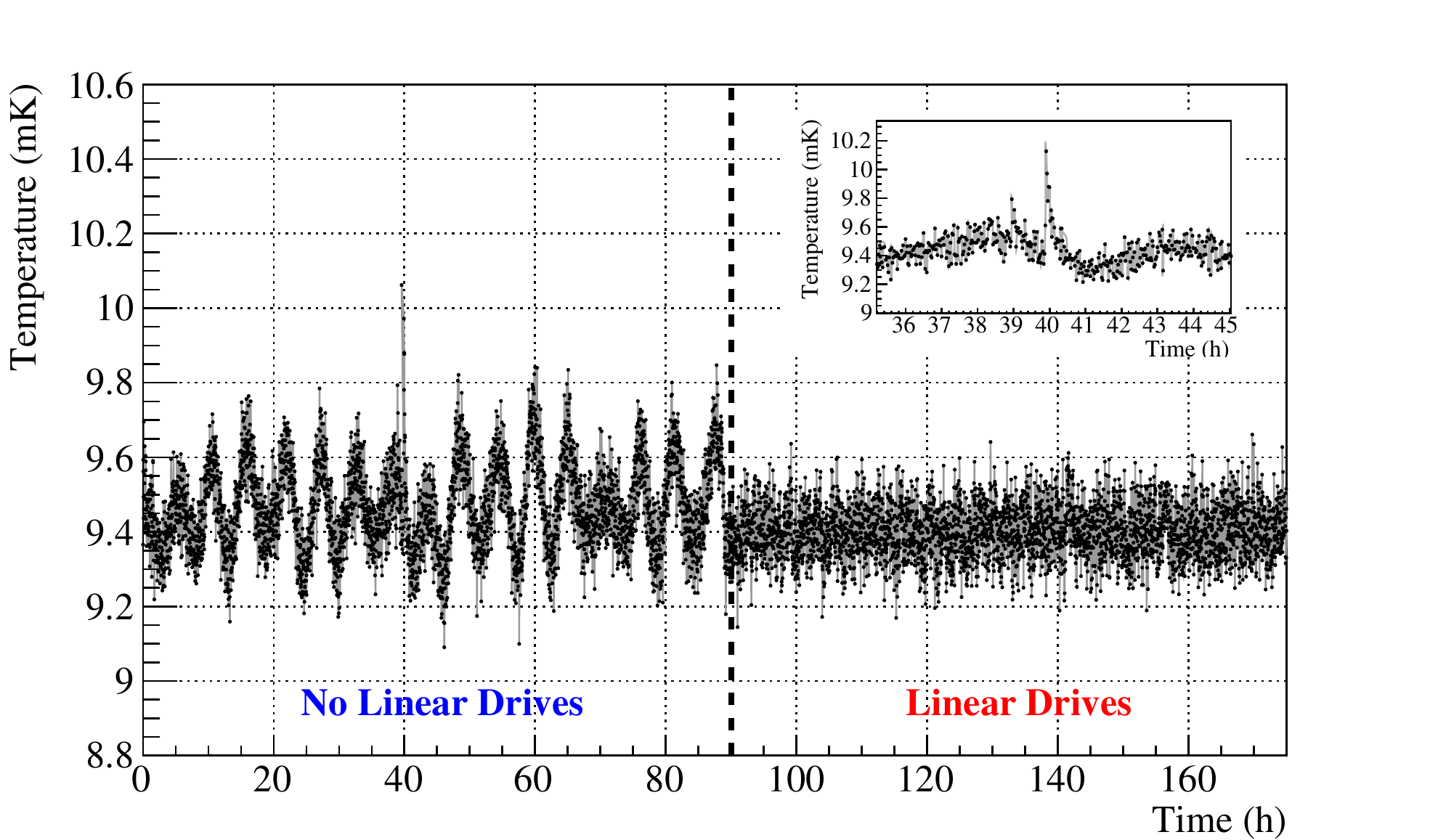} \\[+10pt]
		\caption{Effect of switching the rotary valves feed from Cryomech motors to the LDs. The plot shows the base temperature of the MC plate of the CUORE cryostat, as function of the time (hours from the beginning of the dataset). The temperature is measured by a noise thermometer that acquires 80 1-second samples at 40\,kHz rate. The resulting sampling rate of the plot is approximately 12.2\,mHz, covering a time window of about 7\,days. The insert shows two small earthquakes visible in the data as two spikes approximately 40 hours after the beginning of the data set.}
		\label{fig:LDtemp}
	\end{figure}
   
	In CUORE, each PT rotary valve is controlled by its own LD. We will see in the following that, despite a lower precision in writing the working frequency, the LDs are much less noisy with respect to the Cryomech drives.

\subsubsection{Intrinsic accuracy}

    The intrinsic accuracy of the LD is $\pm$ 5 arcmin. Since we have 25600 step/rev and 21600 arcmin/rev, the accuracy corresponds to $\pm$ 5.926 steps. We operate the LD at the velocity of 0.7 rev/sec, which means 55.8 $\mu$s/step. At this velocity, the LD accuracy in determining the period of the rotations is $\Delta T = \pm$ 330.69 $\mu$s. The expected beat frequency $f_{beat}$ will be of the order of $\Delta$T evaluated at $T = (1.4\,Hz)^{-1}$:
\begin{equation}
    \label{eq:0}
f_{beat} = \frac{[1/T - 1/(T+\Delta T)]}{2} \sim \Delta T / 2T^2
\sim 324\,\mu Hz
\end{equation}

	\begin{figure}[t!]
		\centering
		\includegraphics[width=1.09\columnwidth]{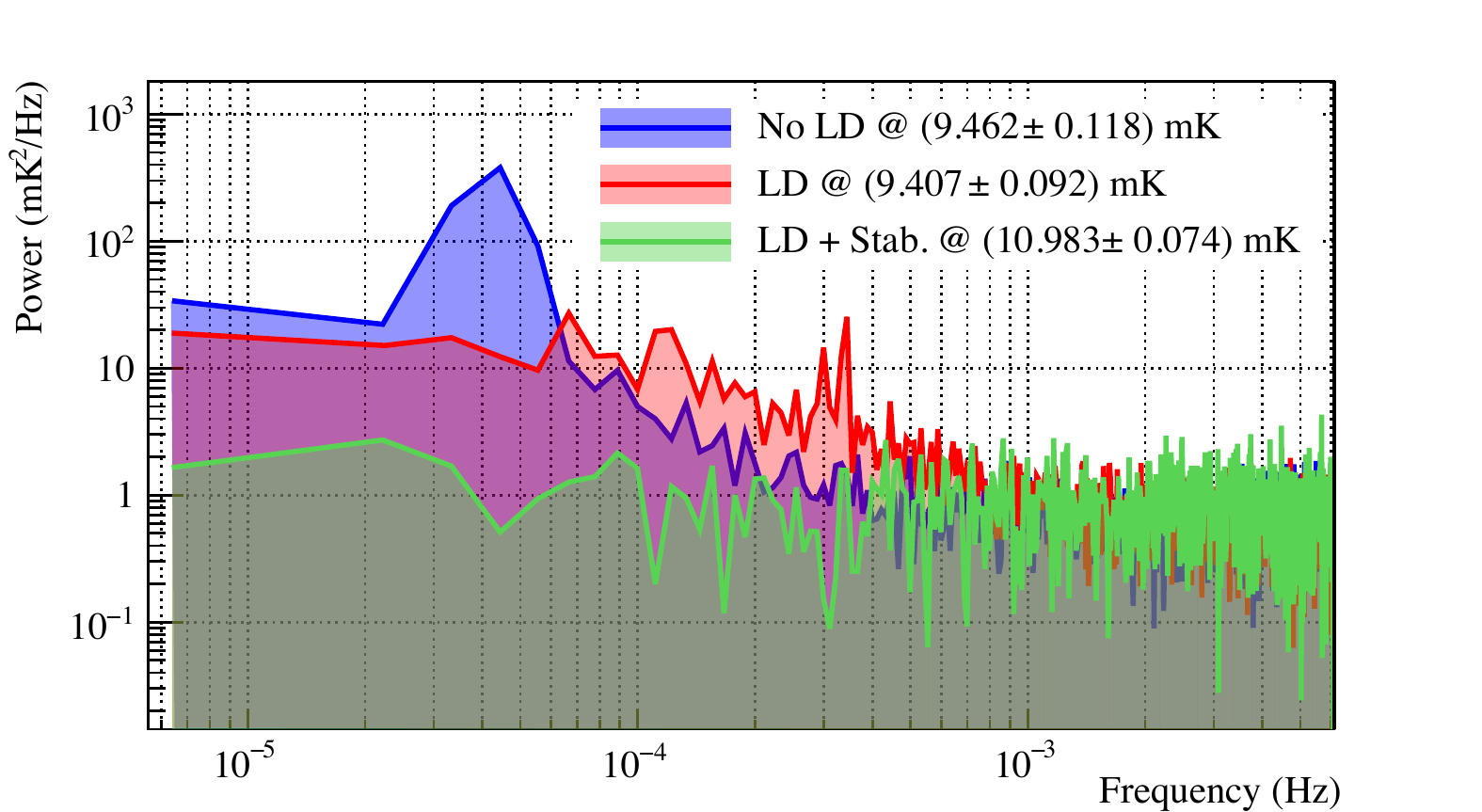} \\[+10pt]
		\caption{Effect of the PT-phase stabilization process on the power spectrum (PS) of the temperature fluctuations on the MC as measured by a noise thermometer. The data have been taken at different temperatures due to \bb ~data taking constraints. The noise has been computed by averaging several 25-hour windows. This shows that the main peak in the blue spectrum at $\sim$45\,$\mu$Hz is due to the 6.2 hours period oscillation in Fig. \ref{fig:LDtemp} before switching the rotary valve feeding to LDs. A similar spectrum has been measured after switching to LDs, without (red) and with (green) PT-phase stabilization. This method strongly attenuates the low frequency peaks and reduce the overall RMS.}
    \label{fig:LDtempFreq}
	\end{figure}

\subsubsection{Driving the rotary valves with LD}

	The replacement of the Cryomech drives with the LDs was implemented only after the CUORE cryostat had reached the base temperature and is shown in Fig. \ref{fig:LDtemp}. Looking at the Power Spectrum (PS) of the temperature fluctuations of the MC as measured by the noise thermometer on the plate (Fig. \ref{fig:LDtempFreq}), an improvement is immediately visible in the noise and stability of the base temperature.
	
	When using Cryomech drives, the MC base temperature as measured by a noise thermometer \footnote{Magnetic Field Fluctuation Thermometer (MFFT-1) from Magnicon.} clearly shows oscillations due to the beating effect from the periodic vibrations of the four PTs, each of which has a slightly different working frequency. The most evident beating shows a period of the order of 6.2 hours, corresponding to $f_{beat} \sim 45\,\mu Hz$ with a noise level of $\sim 400\,mK^2/Hz$. Since this is the highest visible $f_{beat}$ in the noise thermometer PS, and it corresponds to $\Delta f \sim$ 90 $\mu$Hz, we can state that using the Cryomech drives the working frequencies of the PTs are the same within less than 1 part per 10000.
	
	When the PTs are fed by the LDs, the beat frequencies are shifted to higher values. We do see beat frequencies consistent with what expected from Eq. \ref{eq:0}, around 300 $\mu$Hz and 340 $\mu$Hz , with an amplitude that does not exceed $\sim 20\,mK^2/Hz$. The greatest $\Delta$f observed is 680 $\mu$Hz, then we can state that using the LD the working frequencies of the PTs are the same within less than 1 part per 1000.

\subsubsection{Effects and improvements}

	The base temperature PS analysis leads to three conclusions:
\begin{itemize}
\item switching from Cryomech drives to LD does not change the PT working frequencies by more than 1 part per 1000;
\item considering the observed f$_{beat}$, the LD is $\sim$10 times less precise with respect to the Cryomech drive in writing the PT working frequencies;
\item considering the amplitude of the beating peaks, the LD is $\sim$4 times less noisy with respect to the Cryomech drive.
\end{itemize}

As a consequence, we can safely state that the pressure differential between high and low PT inlet should be the same (within 1 part per 1000) after switching from Cryomech drives to LD. Given that the amplitude of the 1.4 Hz oscillation should be proportional to the pressure differential, this means that the vibration level should not change after switching. Any noise improvement obtained with LD does not depend on the pressure differential change, which is negligible. The noise level reduction after switching to LD is then due to the reduced vibrations induced on the motor/rotary valve assembly by the LD, since it offers an high level of stepper discretization dividing the 360$^{\circ}$ angle into 25600 steps.

	When using the LD, the intrinsic noise generated by the valve stepping motor rotation is appreciably reduced. From Fig. \ref{fig:LDtempFreq}, it is clearly visible that LD usage suppresses the beating frequency power amplitude by more than an order of magnitude. Moreover, the overall RMS is reduced by more than 20\%, going from 0.118 mK to 0.092 mK. The reduction of the vibration level has a positive effect on the average MC temperature, decreasing from 9.462 mK to 9.407 mK.
	
	The great improvement seen with the LD, in terms of lower amplitude vibrations, is achieved at the cost of reduced precision in writing the PT working frequency which results in higher beat frequencies. Nonetheless this residual effect is completely washed out by the ``stabilization" algorithm, which is presented in next section.

\subsection{Driving the relative phases}

	Since the detector noise varies with time as the PT relative phases drift, it is plausible that minima of the noise can be obtained for some configuration of the relative phases of the PTs. Finding this configuration and maintaining it would have the double effect of minimizing the noise while making it stable in time.
	
	To prove the feasibility of this approach to active noise cancellation, one needs to show first that it is possible to reproduce a specific relative phase configuration, second that it is possible to identify the most favorable one for noise reduction, and third that the selected phase configuration can be kept constant in time.
	
	Thanks to the capability of controlling its working frequency, the LD enables three crucial operations: identify a phase configuration (calibration); move between two different phase configurations (control) and stabilize a given PT phase configuration in time (stabilization). These three operations are carried out by a real-time custom software as described in the following.

\subsubsection{Calibration}	
    
    A software routine acquires the pressure sensors readings on the compressor low-pressure inlet flexlines of the four active PTs. The periodic pressure signals are sampled at 1\,kHz into 5-seconds windows, then a dedicated function calculates the phase shifts with respect to a reference PT, namely PT$^{(5)}$, applying a cross-correlation algorithm. At the same time, another routine acquires the position of the rotary valves in terms of number of steps issued by the LD. The step number is then linearly mapped to a phase shift. After this conversion -- which we refer to as ``calibration'' -- the rotary valves can be driven directly by specifying a phase offset relative to the reference PT.
    
\subsubsection{Control}
    
    In order to move between two different phase configurations, we implement a PT phase control tool. This algorithm calculates the difference between the current and the desired phase configuration. Then, it applies to each involved PT, a single, small correction (acceleration or deceleration) to desynchronize the current frequencies until they reach the desired phase shifts. The algorithm is designed to never apply to the rotary valves corrections that exceed 80\,mHz, so that the PT working frequency always remains between (1.32, 1.48)\,Hz during the control transient. This means that each phase change could take from hundreds of milliseconds to several seconds, depending on the difference between the initial and final configurations.
     
\subsubsection{Stabilization}

	Driving the rotary valves by LDs still does not guarantee the working frequencies of different PTs to be exactly 1.4 Hz. In order to maintain the selected phase configuration constant in time, we adopt a PT phase stabilization tool that acts as a feedback system. This algorithm applies a series of small corrections (small accelerations or decelerations) to synchronize the working frequency of the selected PT to the one of the reference PT. The corrections on the rotary valve frequency never exceed 20\,mHz, in order to smoothly tune the selected phase configuration.
	
	Through this procedure we are guaranteed to maintain a selected phase configuration. Fig. \ref{fig:LDtempFreq} shows the effect of the LD without (red) and with phase stabilization (green) on the PS of the MC temperature fluctuations of the CUORE cryostat. The stabilization algorithm completely washes out the residual beating peaks from the spectrum and suppresses the overall RMS from 0.092 mK at $\sim$ 9.4 mK to 0.074 at $\sim$ 11.0 mK. Since intrinsic RMS of the noise thermometer is measured in 0.039\,mK and  0.033\,mK at 11\,mK and 9.4\,mK respectively, the actual RMS of the temperature fluctuations of the MC are:
	
	\begin{equation}
	\label{eq:1}
    	\begin{split}
	RMS_{LD} &= \sqrt{(0.092)^2 - (0.033)^2}\,mK \\ &= 0.086\,mK \\
	RMS_{LD+stab} &= \sqrt{(0.074)^2 - (0.039)^2}\,mK \\ &= 0.063\,mK
	\end{split}
	\end{equation}
without and with stabilization, respectively. The uncertainty on the MC heater is negligible \footnote{It has been calculated in 0.5\,$\mu$K and 2.3\,$\mu$K, respectively}, therefore the RMS calculated in Eq. \ref{eq:1} describes the MC temperature fluctuations due to PT mechanical vibration noise. Since the error on the MC temperature scales as $T^{-1}$ \cite{Pobell:2007}, we can extrapolate the RMS without stabilization at 11\,mK to be 0.073\,mK. The stabilization procedure then reduces the fluctuations of the MC temperature by $\sim$14\%.

	\begin{figure}[t!]
	\centering
	\includegraphics[width=1\columnwidth]{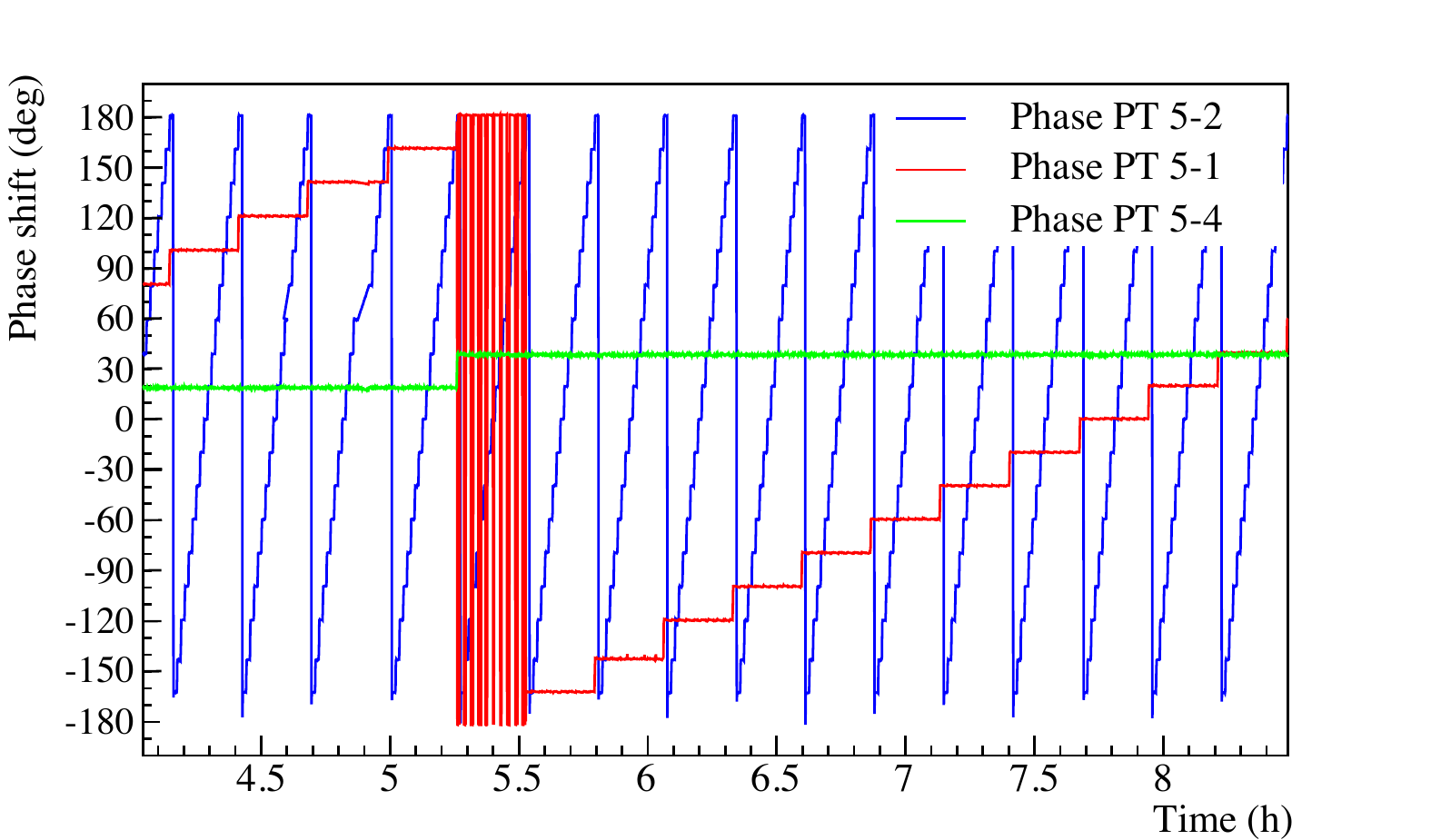} \\[+10pt]
	\caption{Example of how the relative PT phases change during a phase scan. Since the whole phase scan lasts more than three days, we usually divide it into several sub-scans. In this plot the x-axis shows the number of hours from the beginning of a particular phase sub-scan. The vertical red band corresponds to the PT$^{(5)}$-PT$^{(1)}$ phase difference set to -180$^{\circ}$.}
	\label{fig:scan}
\end{figure}

    \subsection{Phase scan}
    
	We can choose any phase configuration (via calibration and control) and make it constant in time (via stabilization). We now have the necessary tools to study different phase configurations searching for the best one.
	
	We implemented an algorithm that automatically scans all the possible phase configurations with a given level of discretization in the parameter space. The software used is a Labview based control interface. It is able to adjust the various PT phase configurations and freeze the system on each of them for a given time, in order to allow the CUORE Data AcQuisition System (DAQ) to acquire the corresponding noise induced on the detector.

    \subsubsection{Process}
	
	There are four active PTs, namely PT$^{(1)}$, PT$^{(2)}$, PT$^{(4)}$ and PT$^{(5)}$. The last of them acts as a reference for the other's phase calculation. We break each of the three phase shift variables (denoted by $x \equiv PT^{(2)}-PT^{(5)}$, $y \equiv PT^{(1)}-PT^{(5)}$, $z \equiv PT^{(4)}-PT^{(5)}$) into steps of 20$^{\circ}$, resulting in $n_p$ = 18$^3= $5832 possible phase configurations.
These are mapped to a scalar running from 0 to 5831, called PhaseID ($ph$), through the following base-18 positional notation:

\begin{equation}
    \scriptstyle
    ph =  \underbrace{\left(\frac{180^{\circ} + x}{20^{\circ}}\right)}_{\equiv x_{18}} \cdot 18^{0} + \underbrace{\left(\frac{180^{\circ} + y}{20^{\circ}}\right)}_{\equiv y_{18}} \cdot 18^{1} + \underbrace{\left(\frac{180^{\circ} + z}{20^{\circ}}\right)}_{\equiv z_{18}} \cdot 18^{2}
\end{equation}
For example, PhaseID $ph$ = 3038 corresponds to the phase configuration (100$^{\circ}$, -60$^{\circ}$, 0$^{\circ}$).

	The scan keeps the system on a constant phase configuration for a given time $t$, taking 5-second detector noise waveforms consecutively. We then move to another phase configuration and repeat the process (Fig. \ref{fig:scan}). The noise waveforms are simultaneously acquired on all the 988 CUORE bolometers. A synchronous digital signal is used to associate a well defined PhaseID to each waveform acquired by the DAQ system. Only waveforms which entirely fit within the same phase configuration are labeled as good waveforms and are associated to a valid PhaseID. 
	
	\begin{table}[b!]
\caption{Duration in hours of the PT phase scan, depending on the number of active PTs ($N$) and on the step size ($S$), assuming to spend $t = $~47\,sec on each phase configuration. Our choice is highlighted.}

\quad

\label{tab:scan}
\rule{0pt}{4.5ex}
\begin{tabular}{l@{\hskip 0.5in} c@{\hskip 0.2in} c@{\hskip 0.2in} c@{\hskip 0.2in} c@{\hskip 0.2in}}\hline \hline
N & \multicolumn{4}{c}{steps [deg]} \\ \cline{1-5} 
\rule{0pt}{2.6ex}  & 45$^{\circ}$  & 36$^{\circ}$ & 30$^{\circ}$ & 20$^{\circ}$ \\ \cline{2-5}
\rule{0pt}{2.6ex}2 & 0.12 & 0.15  & 0.17   & 0.25\\ 
3 & 1 & 1.5  & 2.1  & 4.5   \\
4 & 7.4 & 14.4  & 24.8   & \bf{83.8}   \\
5 & 58.8 & 143.6  & 297.8  & 1507.6   \\
\hline\hline 
\end{tabular}
\end{table}
	
   \subsubsection{Scanning time}
     
	A full scan can be performed in a time given by  
	
	\begin{equation}
	T_{scan} = (d + t) \cdot \underbrace{\left(\frac{360}{S}\right)^{(N-1)}}_{n_{p}}
	\end{equation}
where $t$ is the time spent in measuring the noise on each phase configuration, $S$ is the scan step size expressed in degrees, $N$ is the number of active PTs, and $d \simeq (0.16 \cdot S)$\,sec  is a coefficient taking into account the delay introduced by the algorithm dead times. Notice that $n_{p}$ is the number of possible phase shift configurations, while $(N-1)$ indicates the number of phase shift variables, varying in the range $[$-180$^{\circ}$, 180$^{\circ}]$ and referred to a reference PT. Since the reference PT does not need its working frequency to be ever modified, we choose PT$^{(5)}$ -- which thermalizes $^3$He/$^4$He mixture of one of the injection lines at 40\,K and 4\,K -- as the reference PT. We take $t$ = 47\,sec, in order to acquire up to nine noise samples of 5\,sec each, for each phase configuration.

	T$_{scan}$ is the first, crucial parameter which constrains the applicability of this technique. Since we have enough cooling power to run the CUORE cryostat with only four active PTs out of five (see Sec. \ref{sec:CUORE_PTs} ), we let N = 4. As shown in Tab. \ref{tab:scan}, choosing S = 20$^{\circ}$ we need about 3.5\,days for a complete scan. 

\begin{figure}[t!]
	\centering
    \includegraphics[width=1.0\columnwidth]{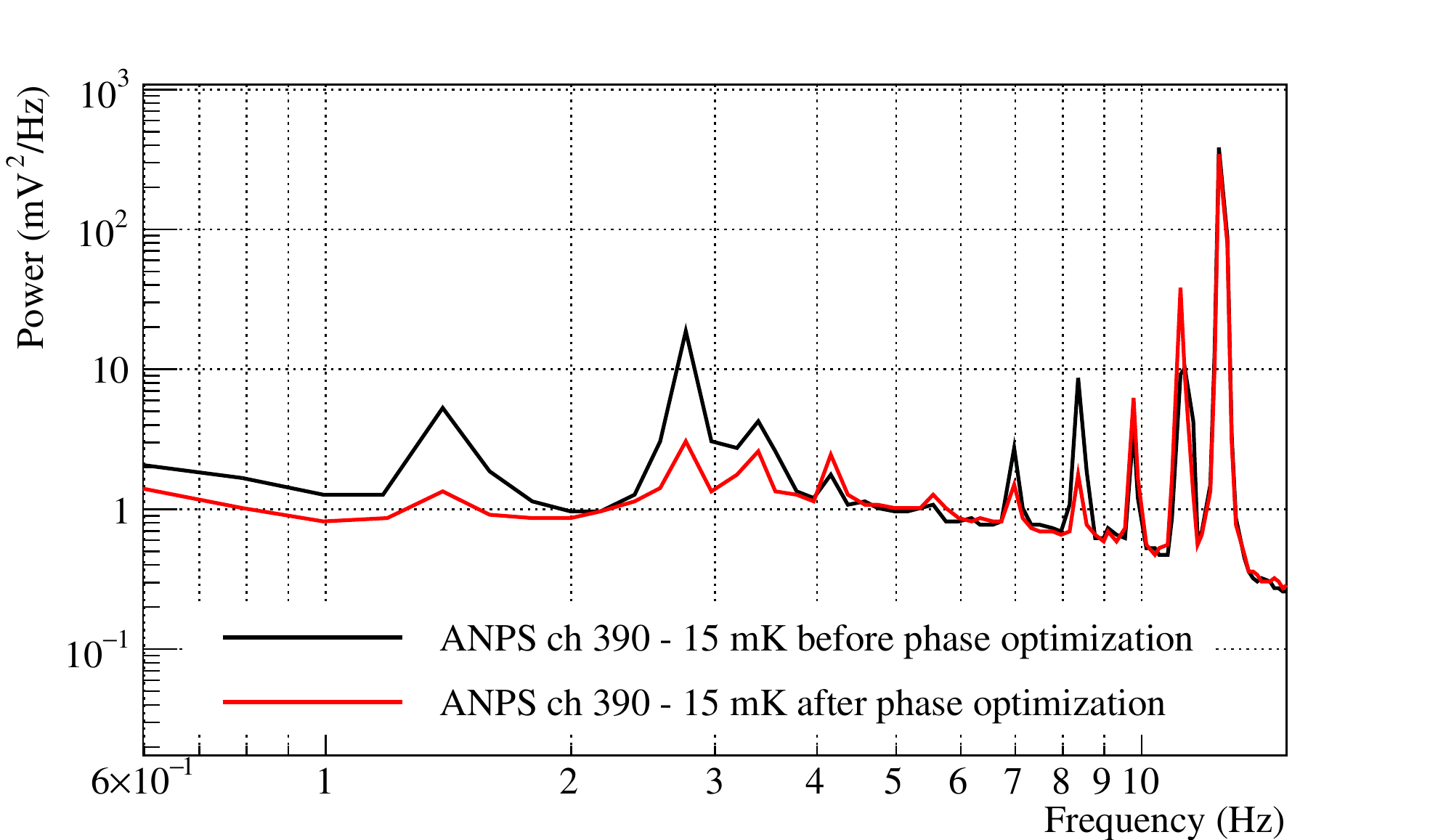}
    \caption{Average Noise Power Spectrum (ANPS) for one of the CUORE bolometer up to 15 Hz. The first ten 1.4 Hz peak harmonics are shown while using the stabilization algorithm for a non-optimized (black) high-noise PT phase configuration. The reduction of the amplitude of the lower harmonics is clearly visible after phase optimization (red, refer to Sec. \ref{sec:APWN}), which keeps the detector in a low-noise phase configuration. The peak around 3.2 Hz is the resonance frequency of the elastomeric mechanical decouplers.}
    \label{fig:ANPS}
\end{figure}

\section{Analysis} \label{sec:anc}
	This analysis proceeds to find an optimal PT-phase configuration that minimizes the detector noise. Since CUORE is a 988-bolometers array, care must be taken to define what a unique ``detector noise'' parameter is in the context of the PT-phase optimization process. In the following sections we describe the analysis method and guidelines for this process.
	
	\subsection{PT harmonics noise}
	
\begin{figure}[t!]
	\centering
    \includegraphics[width=1.0\columnwidth]{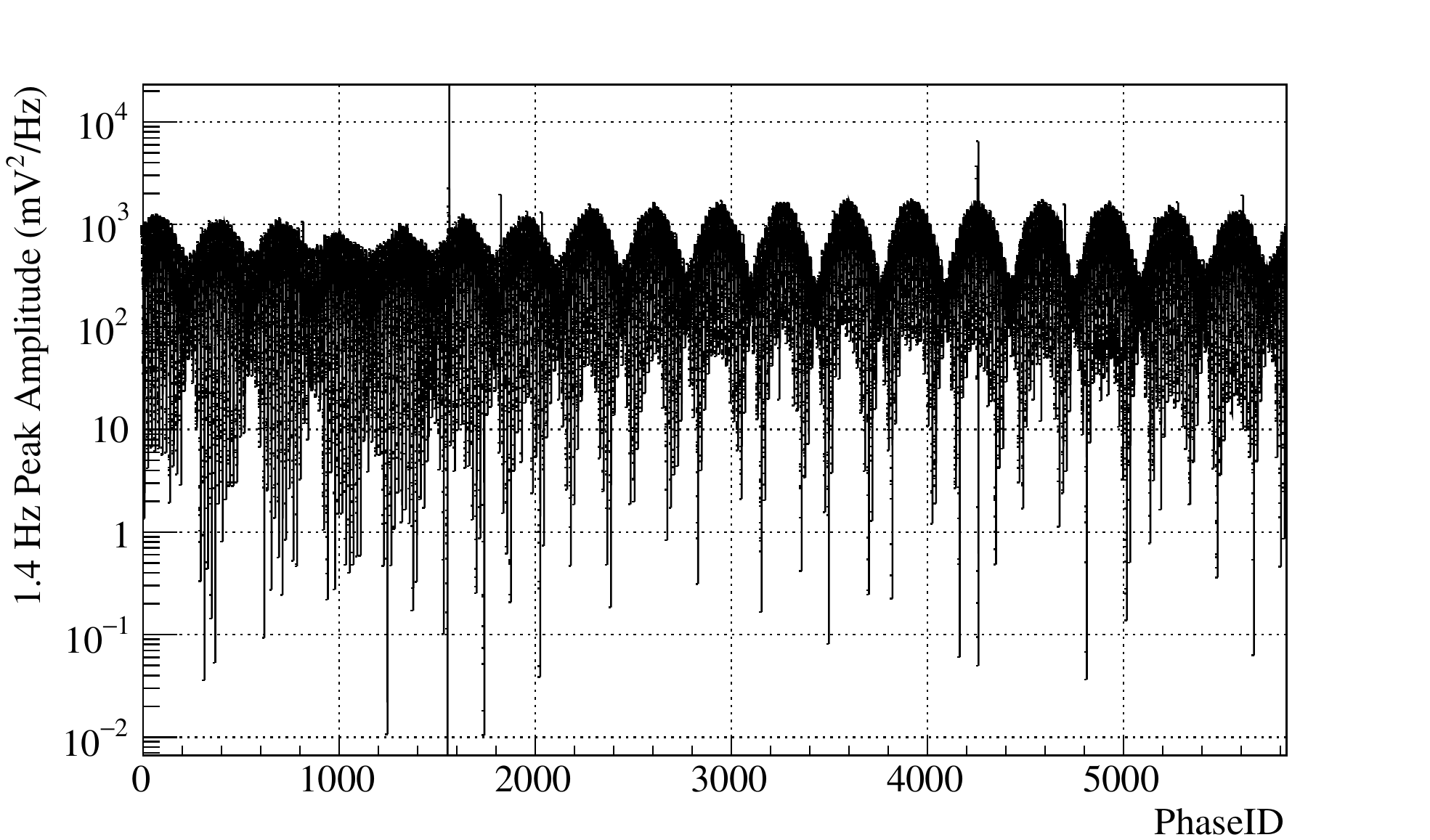}
    \caption{Amplitude of the 1.4\,Hz peak of the NPS for a channel that is particularly sensitive to changes in PT phase configurations. The phase configurations for three PTs are measured relative to a fourth, each representing a 20$^{\circ}$ increment. This results in a 3D phase space with a volume of 18$^3$ unique configurations.}
    \label{fig:PTnoise1}
\end{figure}

	For each channel, the noise contribution due to the PTs can be estimated by examining the intensity of the first ten 1.4\,Hz harmonics

\begin{equation}
\begin{split}
f_n &= f_1, f_2, ..., f_{10} \qquad\textnormal{with}\quad n = 1, ..., 10 \\ &= n \cdot f_0 \qquad\qquad\qquad\qquad f_0 = 1.4\,Hz
\end{split}
\end{equation}
in the Noise Power Spectrum (NPS) estimated for each detector from the acquired waveforms (Fig. \ref{fig:ANPS}). The amplitude of the 1.4\,Hz noise peak and its harmonics is measured for each CUORE bolometer as a function of the PhaseID. As an example, Fig. \ref{fig:PTnoise1} shows the intensity of the 1.4\,Hz peak in the NPS of one CUORE bolometer. It is possible to identify very distinct maxima and minima in the available phase space. A similar behavior is also observed in the higher harmonics of the 1.4\,Hz peak (2.8\,Hz, 4.2\,Hz, 5.6\,Hz, etc.).
   
    \subsection{Average Pulse Weighted Noise} \label{sec:APWN}
	
	Each phase configuration has a dwell time of 47\,sec, collecting up to 9 valid, non-overlapping 5\,sec waveforms for each bolometer. For any given waveform the peak amplitudes of the first 10 PT harmonics are computed (referred to as an NPS amplitude list). Each NPS amplitude list is mapped to the bolometer and phase configuration of the waveform used in its computation. As a result, each channel and phase configuration has a set of $m$ different NPS amplitude lists, where $m$ is the actual number of valid waveforms returned by the DAQ.
	
	Each of the 10 values in an NPS amplitude list, denoted as 
\begin{equation}
\begin{split}
\mathcal{N}^{(i)}_{ph,ch}(f_n) \qquad \textnormal{with} \quad i &= 0, ..., m \leq 9 \\
n &= 0, ..., 10 \\
ph &= 0, ..., 5831 \\
ch &= 1, ..., 988
\end{split}
\end{equation}
is corrected for the front-end gain of the electronics ($\mathcal{G}_{ch}$), and subsequently weighted by the PS of the detector response to a signal event-like mono-energetic energy deposition, referred to as the average pulse ($AP_{ch}(f)$), specific to each detector for a particular working point \cite{Alduino:2016zrl}. This suppresses the frequencies not present in the bolometer signal rather than assigning a uniform weight to all frequencies. As the bolometric signal is dominated by frequencies below 3\,Hz, the importance of the lower harmonics of the PTs (1.4\,Hz, 2.8\,Hz) is largely enhanced with respect to the higher ones.

	Subsequently each of these AP-weighted NPS amplitude lists is summed over its 10 values to obtain a single value representing the AP-weighted 1.4\,Hz response: 
\begin{equation}
\begin{split}
\mathcal{N}^{(i)}_{ph,ch} = \sum^{10}_{n = 1} AP_{ch}(f_n) \cdot \mathcal{G}^{-1}_{ch} \cdot \mathcal{N}^{(i)}_{ph,ch}(f_n)
\end{split}
\end{equation}

	We denote any specific such AP-weighted noise value as: $\mathcal{N}^{(i)}_{ph,ch}$, where $ch$ is the CUORE bolometer channel (1-988), $ph$ is the PT scan PhaseID (0-5831), and $i$ ranges from 1 to $m$, being $m$ the number of valid waveforms for each ($ph$, $ch$) pair.
	
	After the AP-weighted sum process is completed, the median of the $m$ $\mathcal{N}^{(i)}_{ph,ch}$ values for each ($ph, ch$) pair is taken to obtain the typical AP-weighted summed noise for each channel and for each phase configuration\footnote{The median is used to avoid sensitivity to outliers.}:
\begin{equation}
\mathcal{N}_{ph,ch} = \textnormal{median}\left(\mathcal{N}^{(i)}_{ph,ch}; i\right)
\end{equation}
This is referred to as unnormalized noise.

\begin{figure}[t]
		\centering
		\includegraphics[width=1.\columnwidth]{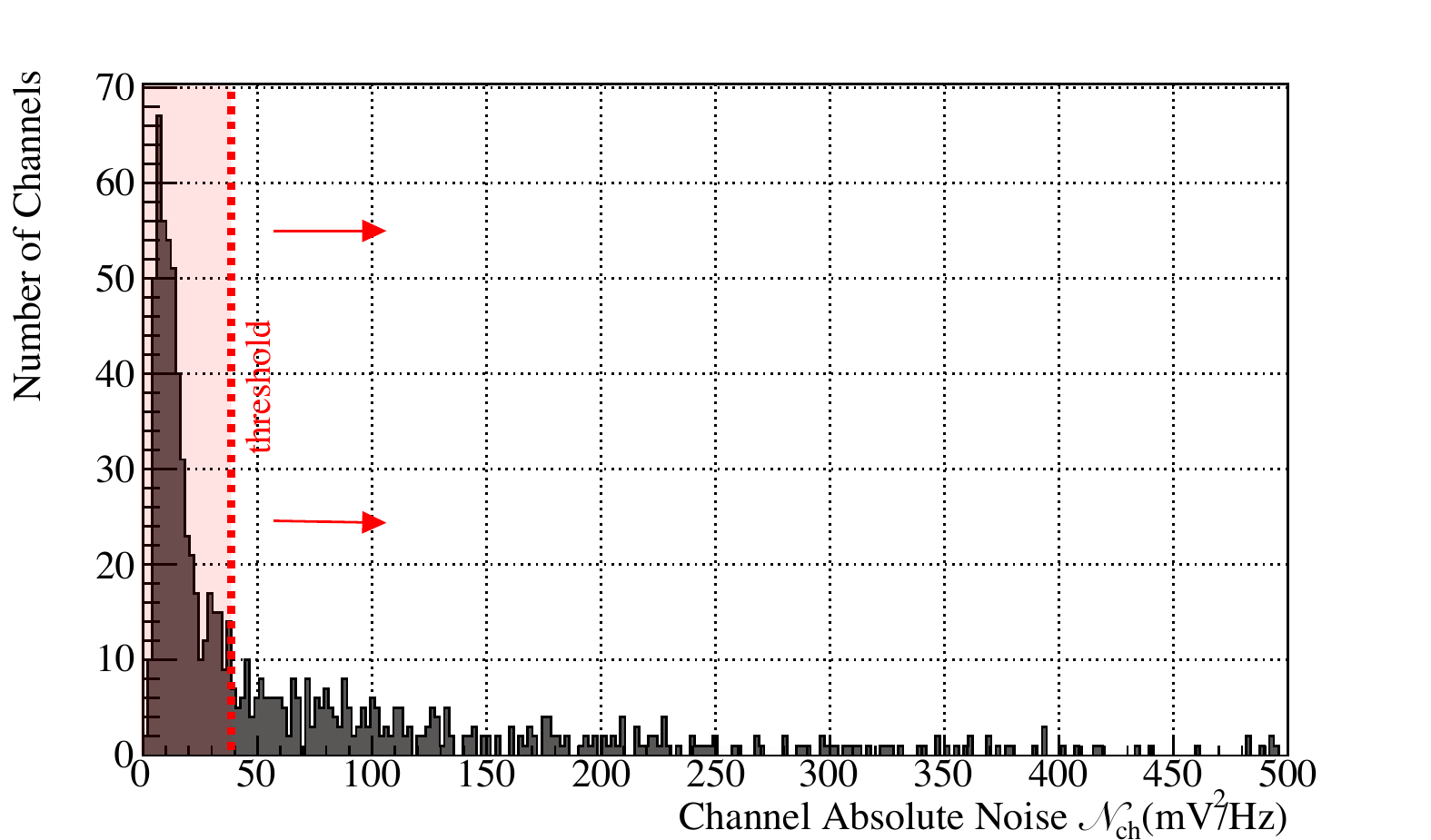} \\[+10pt]
        \caption{Histogram of the total absolute noise $\mathcal{N}_{ch}$. For a specific channel, the median across all phaseIDs of $\mathcal{N}_{ph,ch}$ is taken. During the optimization process, only the channel above threshold (vertical red line, see text for details) are considered.}
		\label{fig:PTnoise_abs_noise}
	\end{figure}

	\subsubsection{Channel absolute noise}
    \label{sec:abs_noise}

	For a given $ch$, the median of the $\mathcal{N}_{ph,ch}$ distribution across all $ph$, namely $\mathcal{N}_{ch}$, is taken to create a set of typical channel noise values:
\begin{equation}
\mathcal{N}_{ch} = \textnormal{median}\left(\mathcal{N}_{ph,ch}; ph\right)
\end{equation}
This is referred to as the channel absolute noise. This  variable quantifies the impact of PT vibrations on the NPS of a particular channel. Higher $\mathcal{N}_{ch}$ will correspond to channels more influenced by PTs-induced noise.

\begin{figure*}[!t]
\begin{center}$
\begin{array}{c}  
\subfigure[]{\includegraphics[width=1.0\linewidth]{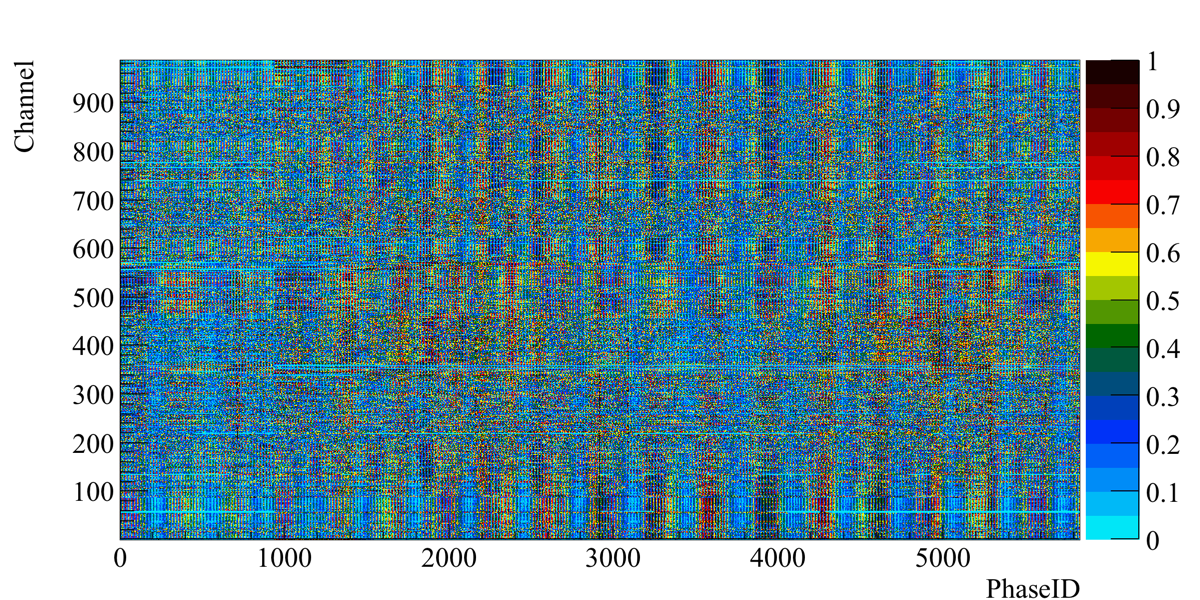}}\label{fig:PTnoise_3D} \\
\subfigure[]{\includegraphics[width=0.98\linewidth]{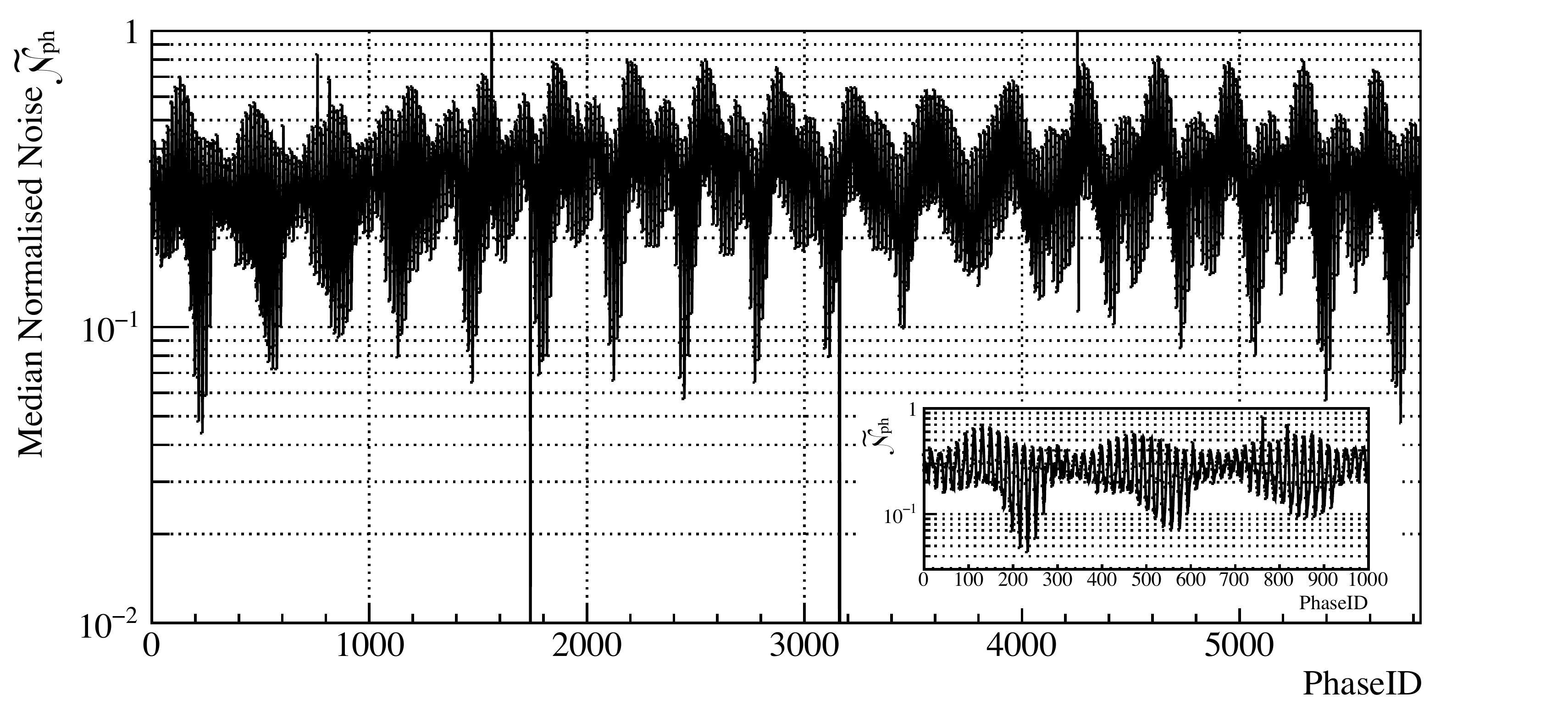}}\label{fig:PTnoise_2D}  
\end{array}$
\end{center}
\caption{(a)~Normalized total AP-weighted sum of the 1.4\,Hz harmonics noise, $\widetilde{\mathcal{N}}_{ph,ch}$. Each channel is itself normalized across all PhaseID values, trimming away the lower 2\% and upper 2\% of $\widetilde{\mathcal{N}}_{ph,ch}$ values. Each cell contains the normalized value of the noise. (b)~The median across all channels of $\widetilde{\mathcal{N}}_{ph,ch}$ is taken to give the normalized noise as a function of PhaseID, $\widetilde{\mathcal{N}}_{ph}$. The 2 bands that extend out of range are due to periods of time where an earthquake occurred and are excluded from the optimization procedure. The inset shows a zoom of the region where the minimum occurs.} 
\label{fig:norm_noise_entire}
\end{figure*}

	Generally, the behavior of the detector appears to be different on a channel by channel basis. The majority of the channels are distributed at low noise, with a tail of channels exhibiting up to two orders of magnitude higher absolute noise overall (see Fig. \ref{fig:PTnoise_abs_noise}).

	\subsubsection{Normalized noise}

	For some detector channels which are weakly susceptible to PT-induced noise, there is the risk that small variation around a potential $ph$ minimum in the unnormalized noise $\mathcal{N}_{ph,ch}$ could be hidden by the large overall variation of the channels which are strongly susceptible to  PT-induced noise.

	A normalized quantity, $\widetilde{\mathcal{N}}_{ph,ch}$, can be obtained by taking for each channel a subset of $ph$ values which do not belong to the lower 2\% and upper 2\% of the $\mathcal{N}_{ph,ch = fixed}$ distribution and normalizing this to the range [0,1]\footnote{Phase configurations that are trimmed away would either exhibit a negative or greater than 1 $\widetilde{\mathcal{N}}_{ph,ch}$ value.}. This quantity -- referred to as normalized noise -- is shown in Fig. \ref{fig:norm_noise_entire} (a). Changes in detector noise that are dependent upon phase configuration $ph$ are noticeable after normalization in the form of vertical banding of high and low values.

\begin{figure} [t!]
\begin{center}$
\begin{array}{c}  
\subfigure[]{\includegraphics[width=1.0\linewidth]{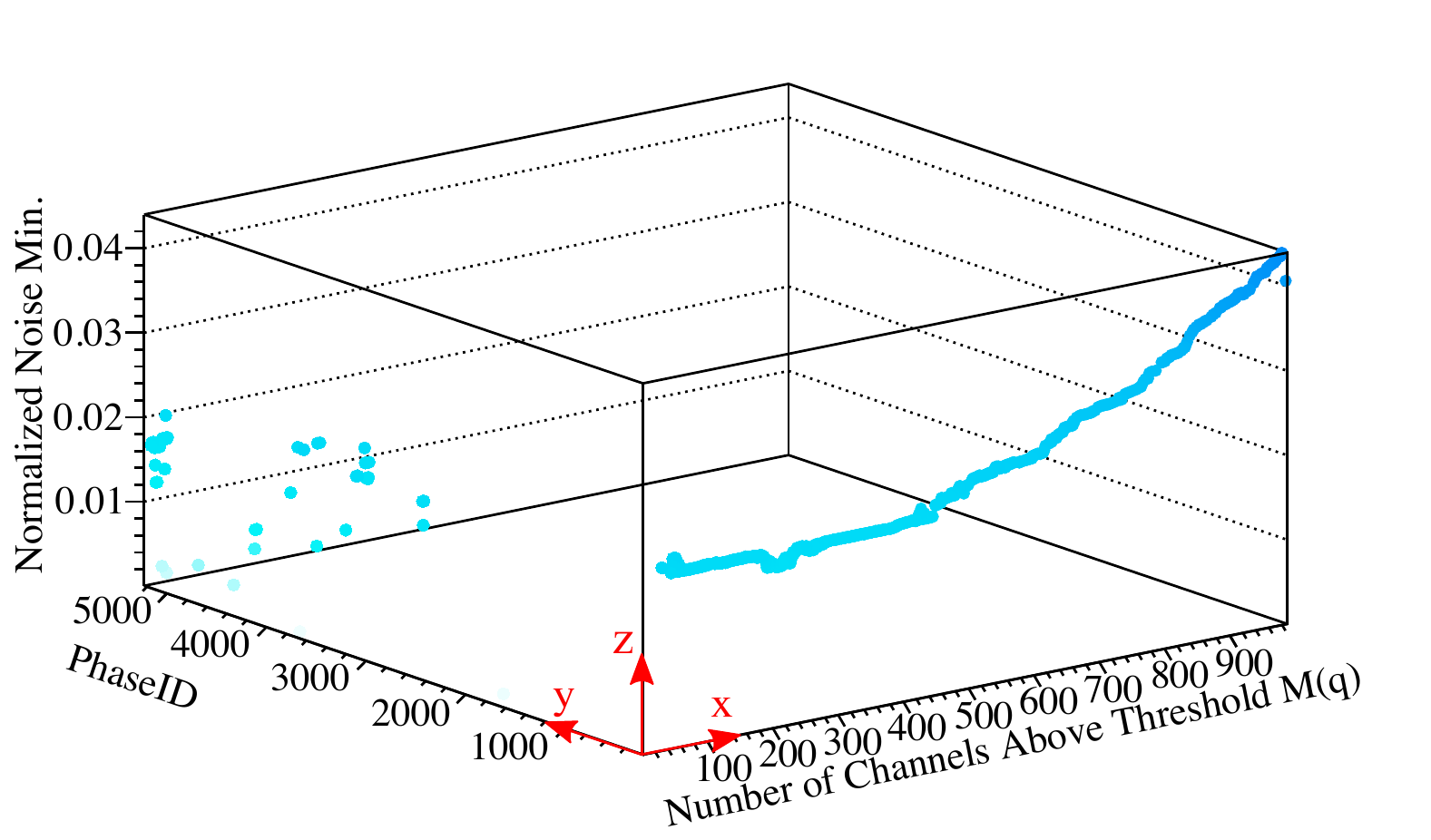}}\label{fig:opt_ph3D} \\
\subfigure[] {\includegraphics[width=1.0\linewidth]{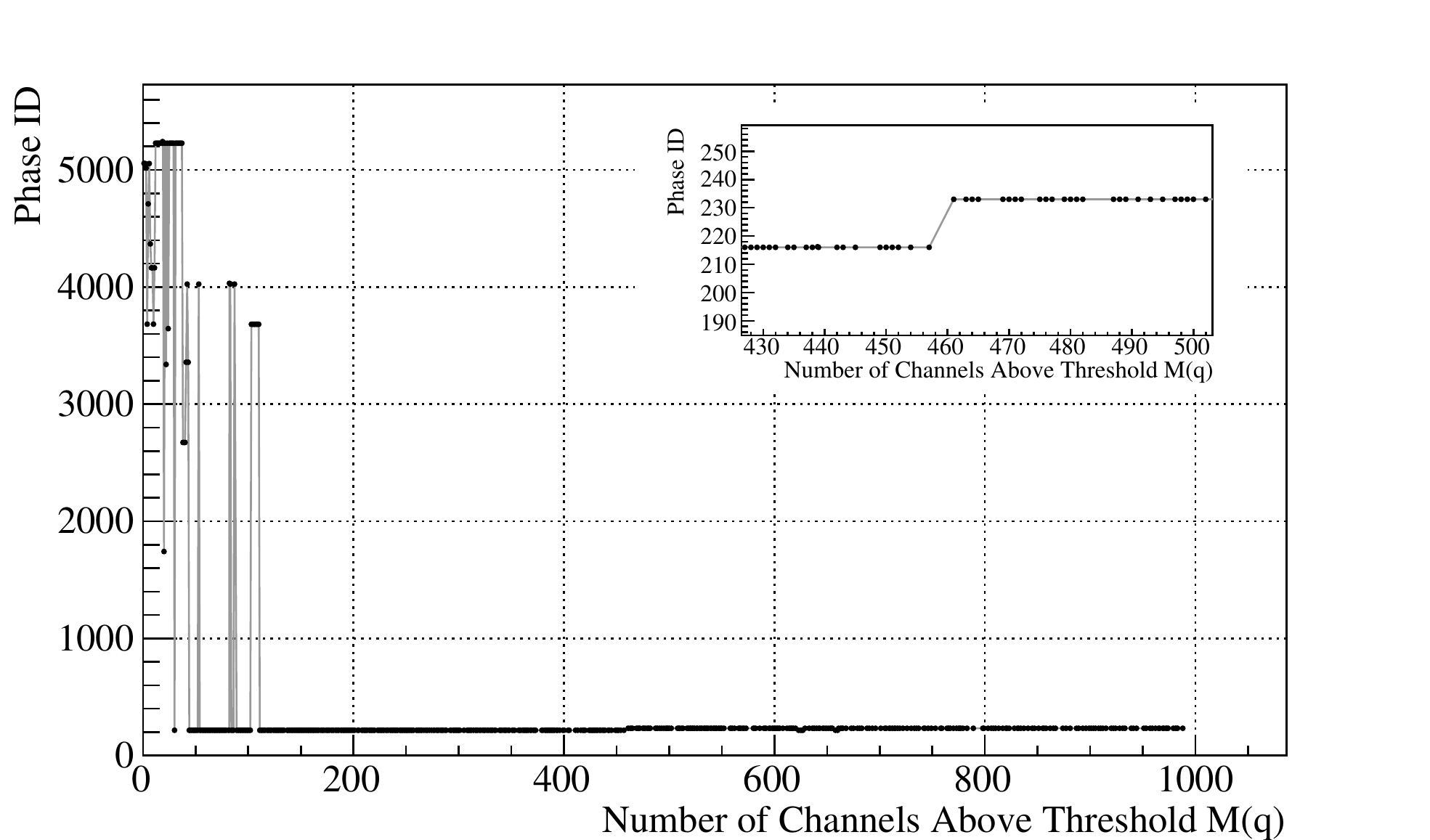}}\label{fig:opt_ph2D}  
\end{array}$
\end{center}
\caption{Optimal PhaseID identification. (a)~Each selected threshold $q$ corresponds to a point in the plot, identified by a subset of channels above threshold (x-axis) whose minimum normalized noise (z-axis) occurs at a particular phase configuration (y-axis). Points are colored by the minimum normalized noise value using the same color scale of Fig. \ref{fig:norm_noise_entire}(a). (b)~Projection of the previous plot onto the x-y plane. When considering more than $\sim$ 10\% of the detector, only two phase configurations come out as optimal: $ph =$216 and $ph =$233. The inset zooms on the region where the change happens.} \label{fig:opt_ph}
\end{figure}

\subsection{Optimal phase configuration}
\label{sec:criteria}

	The normalized noise $\widetilde{\mathcal{N}}_{ph,ch}$ allows comparison of channels in terms of their noise relative variations, regardless of having very different absolute noise levels. For a particular $ph$ value, the distribution of normalized noise across all channels is examined and the median
	
\begin{equation}
\widetilde{\mathcal{N}}_{ph} = \textnormal{median}\left(\widetilde{\mathcal{N}}_{ph, ch}; ch\right)
\end{equation}
is taken to represent the detector's typical normalized noise response to this particular phase configuration (Fig. \ref{fig:norm_noise_entire}(b)). Then the $ph$ which returns the minimum $\widetilde{\mathcal{N}}_{ph}$ value is taken as the best phase configuration.

\subsubsection{Threshold}

	In order not to be biased by a group of channels weakly affected by PT noise, care must be taken in selecting the channels on which the optimization process shall run. To iteratively remove channels from this computation, a threshold $q$ is applied on the variable $\mathcal{N}_{ch}$ introduced in Sec. \ref{sec:abs_noise} (see Fig. \ref{fig:PTnoise_abs_noise}). Once a value of threshold $q$ is fixed, a channel is not considered for the current iteration if $\mathcal{N}_{ch} < q$. Thus for a particular threshold $q$, the subset of $M(q) < 988$ detector normalized noise responses, $\widetilde{\mathcal{N}}_{ph}\left(q\right)$, is what is used for optimization. The optimization process then trims away the lowest 0.1\% of the $\widetilde{\mathcal{N}}_{ph}\left(q\right)$ distribution and searches for the minimum of the remaining values. From here, the value of $ph$ where the minimum occurs is identified as the best phase configuration for the particular subset of $M(q) < M(q=0) = 988$ channels whose absolute noise $\mathcal{N}_{ch}$ exceeds the threshold $q$.

	The very first iteration of the optimization sets $q = 0$ and is indeed applied to the whole detector. The process is then repeated on progressively increasing thresholds and iterated until all channels are excluded by the threshold.

\section{Results and discussion} \label{sec:results}
	\begin{table}[t!]
\caption{Optimal PT phase configurations}
\label{tab:opt_ph}
\rule{0pt}{3.5ex}
\begin{tabular}{l@{\hskip 0.2in} c@{\hskip 0.2in} c@{\hskip 0.2in} c@{\hskip 0.2in}}\hline \hline
PhaseID & \multicolumn{3}{c}{phase shift [deg]} \\ \cline{1-4} 
\rule{0pt}{2.6ex}  & PT$^{(2)}$-PT$^{(5)}$  & PT$^{(1)}$-PT$^{(5)}$ & PT$^{(4)}$-PT$^{(5)}$ \\ \cline{2-4}
\rule{0pt}{2.6ex}216 & -180$^{\circ}$    & +60$^{\circ}$   & -180$^{\circ}$\\ 
233 & +160$^{\circ}$   & +60$^{\circ}$  & -180$^{\circ}$   \\
\hline\hline 
  \end{tabular}
\end{table}
	
	\begin{figure}[b!]
		\centering
		\includegraphics[width=1.0\columnwidth]{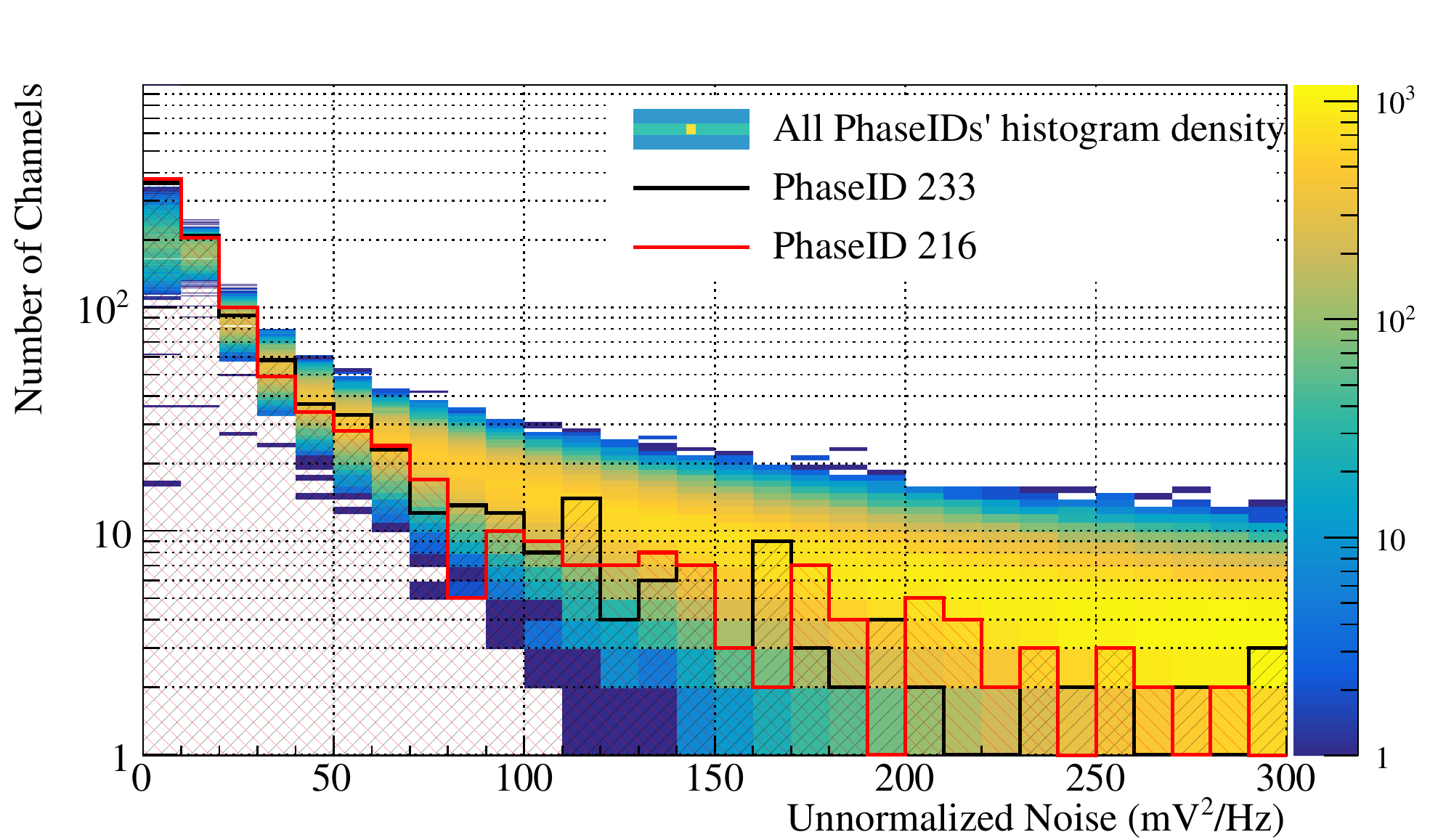} \\[+10pt]
        \caption{This figure shows a density histogram of the median noise across all channels for each individual phase configuration binned by the typical AP-weighted noise value. The colored z-axis shows the number of PhaseID histograms per pixels. We see that a majority of phase configurations do have a few hundred channels with low noise but that there is a long tail typically to the fall off of noise. In contrast the two best performing PhaseIDs (233 in black and 216 in red) are at the top of the profile at the very lowest values and fall quickly to the lower edge of the profile as the noise bins increase in value. This is the response one might expect for configurations that minimize noise.}
	\label{fig:hyst}
	\end{figure}  
	
	Fig. \ref{fig:opt_ph} shows the results from this analysis. In the origin of the x-axis of the 3D plot in Fig. \ref{fig:opt_ph}(a), the threshold $q$ is higher than the highest $\mathcal{N}_{ch}$. While $q$ decreases, more channels are included into the computing of the optimal $ph$. Its value is more clearly visible from Fig. \ref{fig:opt_ph}(b). When considering a number of channels in the range 100-457, $ph$ = 216 is favored; including a number of channels higher than 457, $ph$ = 233 is mostly preferred. These two optimal PhaseIDs found by the analysis procedure translate in the real PT phase configurations shown in Tab. \ref{tab:opt_ph}. As shown by Fig.\ref{fig:300K}, the active PTs are coupled two by two in term of their positioning on top of the CUORE cryostat. The best relative phase configurations indicates that at the optimal working PT$^{(2)}$ and PT$^{(4)}$ are in phase, operating in opposition of phase with respect to PT$^{(5)}$. Furthermore, PT$^{(1)}$'s best phase occurs when it operates approximately in quadrature phase with respect to its nearest partner, PT$^{(2)}$. In the best phase configuration, the active PTs cause destructive interference of the vibrations injected into the cryostat. Fig. \ref{fig:ANPS} shows an example of the lowering of the PT peaks in the NPS of one of the CUORE channels using $ph$ = 233. Fig. \ref{fig:hyst} shows the histograms of the PT absolute noise distribution for the  two optimal PhaseIDs.
Overlayed to these, the 2D distribution of phase configurations as a function of the number of channels and of the absolute noise is shown as a color map.
	
	The PhaseID 216 (red) and 233 (black) histograms clearly lie out of the average trend, having the highest number of channels with lower noise and reducing the amount of channels with higher noise.

\section{Conclusions} \label{sec:conclusions}
	We developed a technique to mitigate the effect of mechanical vibrations injected into a cryogen-free refrigerator by the simultaneous operation of more than one Pulse Tube cryocooler. We demonstrated that driving the PT rotary valves by means of external microstepping control devices intrinsically reduces the amount of vibration injected into the cryogenic system. Moreover, this allows the online change and stabilization of the phases of their driving signals, thereby providing an efficient handle to search for the phase configuration which minimizes the overall detector noise.
	
	We implemented a PT phase scan technique and validated the stability and reliability performance of the control software. Furthermore, we developed a robust analysis method for finding minima and define an optimal working phase configuration.
	
	This technique shows how the PT-induced noise can be suppressed on cryostats needing more than one PT cryocooler to be operated. Provided that the LDs have sufficient dynamic range, this method can be extended to any type of PT, such as HF-PT \cite{Dietrich:2010} or those driven by piston compressors \cite{Houlei:2012}.
	
	The active noise cancellation technique described in this paper is being applied to the CUORE PT cryocoolers during the data taking, whose first official data release can be found in \cite{Alduino:2018}.
	
	By addressing a long standing impediment to growth in the field, this method demonstrates that low temperature detector experiments are scalable to very large cryogen-free cryogenic infrastructures, opening the path for rare event searches with ton-scale bolometric detectors operated by cryogen-free refrigerators.

\quad

\section{Acknowledgments} \label{sec:ack}
% ---------------------------------------
The authors thanks the CUORE Collaboration, the directors and staff of the Laboratori Nazionali del Gran Sasso and the technical staff of our laboratories. In particular, we would like to thank Dr. Luigi Cappelli for his contribution to the installation of the software interface. We would also like to thank Prof. Giorgio Frossati, Dr. Alain Benoit, Dr. Alessandro Monfardini and Dr. Luca Taffarello for the useful discussions. This work was supported by the Istituto Nazionale di Fisica Nucleare (INFN); the European Research Council (FP7/2007-2013) under Contract CALDER No. 335359; the Director, Office of Science, of the U.S. Department of Energy under Contract Nos. DE-AC02-05CH11231, DE-AC52-07NA27344, and DE-SC0012654; the DOE Office of Nuclear Physics under Contract Nos. DE-FG02-08ER41551, and DEFG03-00ER41138; the National Science Foundation under Grant Nos. NSF-PHY-0605119, NSF-PHY-0500337, NSF-PHY-0855314, NSF-PHY-0902171, NSF-PHY-0969852, NSF-PHY-1307204, NSF-PHY-1314881, NSF-PHY-1401832, and NSF-PHY-1404205.

\bibliographystyle{apsrev4-1}
	\bibliography{ref_arXiv,ref_Books,ref_Others,ref_Papers,ref_Proceedings,ref_Talks,ref_Theses,ref_web}

\end{document}